\documentclass[letterpaper]{article} 
\usepackage{aaai23}  
\usepackage{times}  
\usepackage{helvet}  
\usepackage{courier}  
\usepackage[hyphens]{url}  
\usepackage{graphicx} 
\usepackage{natbib}  
\usepackage{caption} 
\usepackage{algorithm}
\usepackage{algorithmic}

\usepackage{dblfloatfix}
\usepackage{newfloat}
\usepackage{listings}
\DeclareCaptionStyle{ruled}{labelfont=normalfont,labelsep=colon,strut=off} 
\floatstyle{ruled}
\newfloat{listing}{tb}{lst}{}
\floatname{listing}{Listing}
\usepackage{xcolor}

\renewcommand*{\thefootnote}{\fnsymbol{footnote}}

\title{Taxonomizing and Measuring Representational Harms: A Look at Image Tagging}
\author {
    Jared Katzman\footnotemark[1]\textsuperscript{\rm 1},
    Angelina Wang\footnotemark[1]\textsuperscript{\rm 2},
    Morgan Scheuerman\textsuperscript{\rm 3},
    Su Lin Blodgett\textsuperscript{\rm 4},
    Kristen Laird\textsuperscript{\rm 5},
    Hanna Wallach\textsuperscript{\rm 4},
    Solon Barocas\textsuperscript{\rm 4}
}
\affiliations {
    \textsuperscript{\rm 1} University of Michigan\\
    \textsuperscript{\rm 2} Princeton University\\
    \textsuperscript{\rm 3} University of Colorado, Boulder\\
    \textsuperscript{\rm 4} Microsoft Research\\
    \textsuperscript{\rm 5} Microsoft\\
}

\usepackage{bibentry}

\begin{document}

\maketitle

\begin{abstract}
In this paper, we examine computational approaches for measuring the ``fairness'' of image tagging systems, finding that they cluster into five distinct categories, each with its own analytic foundation. We also identify a range of normative concerns that are often collapsed under the terms ``unfairness,'' ``bias,'' or even ``discrimination'' when discussing problematic cases of image tagging. Specifically, we identify four types of representational harms that can be caused by image tagging systems, providing concrete examples of each. We then consider how different computational measurement approaches map to each of these types, demonstrating that there is not a one-to-one mapping. Our findings emphasize that no single measurement approach will be definitive and that it is not possible to infer from the use of a particular measurement approach which type of harm was intended to be measured. Lastly, equipped with this more granular understanding of the types of representational harms that can be caused by image tagging systems, we show that attempts to mitigate some of these types of harms may be in tension with one another.
\end{abstract}

\section{Introduction}
\footnotetext[1]{Authors contributed equally to this work}
\renewcommand*{\thefootnote}{\arabic{footnote}}
\setcounter{footnote}{0}

In recent years, computer vision has gone from a largely academic research topic to a set of technologies that touch the lives of people in a variety of impactful ways. With this success comes the importance of understanding when computer vision systems can cause harms. In this paper, we consider some of the harms that can be caused by a seemingly mundane, yet widespread, computer vision task: image tagging.

Prior work has already documented several problematic cases of image tagging. For example, researchers have identified harms caused by gender classification systems and other systems that sort people into social groups based on their appearance \cite{keyes2018misgendering, barlas2021see}, demonstrated that the performance of image tagging systems can vary according to the identities of the people depicted in images \cite{hendricks2018women, zhao2017men}, and revealed a variety of different ways that such systems can perpetuate stereotypes \cite{schwemmer2020diagnosing, bhargava2019exposing}. These findings have been held up as evidence that image tagging systems can be ``unfair,'' ``biased,'' or even ``discriminatory.'' However, in many of these cases, it can be difficult to identify the full range of normative values at stake.\looseness=-1

We aim to bring greater structure and precision to discussions of some of the harms caused by image tagging systems and the many computational approaches that have been proposed to measure them. We first provide relevant background by defining image tagging and revisiting the distinction between allocational and representational harms~\cite{barocas2017problem}---emphasizing that problematic cases of image tagging can 
affect the understandings, beliefs, and attitudes that people hold about particular social groups. We then analyze previously proposed computational approaches for measuring the ``fairness'' of image tagging systems, clustering them into five distinct categories, each with its own analytic foundation.\footnote{We restrict our focus to  only quantitative measurement approaches, setting aside a survey of qualitative measurement approaches \citep[e.g.,][]{barlas2019makes, hu2018crowd, 10.1145/2702123.2702520, miceli2020power, 10.1145/3392866, offert2020metapictures} for future work.\looseness=-1} Next, we identify four types of representational harms that can be caused by image tagging systems, moving beyond generic concerns about ``unfairness,'' ``bias,'' or ``discrimination.'' We consider how different computational measurement approaches map to each of these types of harms, finding that each approach can be used to measure all four types. In light of this, we argue that no single measurement approach is definitive and that, without explicit reasoning, it is not possible to infer from the use of a particular measurement approach which type of harm was intended to be measured.
Lastly, we conclude by exploring tensions between attempts to mitigate different types of representational harms---and what these tensions~might mean for developing appropriate mitigations.\looseness=-1

\section{Background}

In this section, we define both image tagging and 
representational (versus allocational) harms~\cite{barocas2017problem}.

\subsection{Image Tagging}
Image tagging is the task of applying tags to an image so as to describe salient aspects of its contents. Image tagging differs from the related task of object recognition, where the goal is to identify all objects that are depicted in an image. Crucially, although both tasks involve applying tags to images, object recognition focuses solely on objects and is concerned with completeness, whereas image tagging focuses on other properties (e.g., actions, facial expressions, social groups) as well as objects and is concerned with salience. 

The two tasks also differ in terms of the ways the resulting tags are typically used. The tags applied by image tagging systems are often intended for human consumption. For example, image tagging systems are commonly used to generate alt text descriptions of images for blind and low-vision users \cite{wu2017fb}, provide search or grouping functionality in photo-management software, and enable users of search engines to find images based on their contents as well as their accompanying text. In contrast, the tags applied by object recognition systems are often used to automate tasks, obviating the need for human involvement. For example, semi-autonomous vehicles rely on object recognition systems to detect obstacles, but may take autonomous actions to avoid colliding with those obstacles rather than simply communicating their presence to drivers.\looseness=-1

Image tagging has a unique normative valence because it is concerned with salience and applies tags that are often intended for human consumption. The tags applied to images by image tagging systems are expressions of what---and who---is important and how they deserve to be described. Because the tags are consumed by people, they affect what people view as salient and how they understand the world.\looseness=-1

\subsection{Representational Harms}

Much of the foundational work on the ``fairness'' of machine learning systems has focused on what \citet{barocas2017problem} call allocational harms---that is, when people belonging to particular social groups are unfairly deprived of access to important opportunities or resources. Allocational harms often arise in domains like employment, finance, and housing. However, computer vision systems, like natural language processing systems, are not typically used to allocate opportunities or resources and are therefore less prone to causing allocational harms \cite{barocas2017problem, crawford2017trouble, blodgett2020language}. Instead, these systems produce outputs that can affect the understandings, beliefs, and attitudes that people hold about particular social groups, and thus the standings of those groups within society.  \citet{barocas2017problem} call these representational harms.
Concerns about representational harms have motivated research examining, for example, the ways that search engines represent some social groups in less favorable ways than they do others (e.g., returning more images of men than women in response to search queries for ``doctor'' \cite{kay2015unequal} or returning pornographic websites in response to queries for ``Black girls'' but not ``white girls'' \cite{noble2018algorithms}). 

\section{Computational Measurement Approaches}
\label{sec:survey}
Equipped with definitions of image tagging and representational harms, we now analyze previously proposed computational approaches for measuring the ``fairness'' of image tagging systems. We focus on prior work that self-identifies as measuring ``unfairness,'' ``bias,'' or ``discrimination.'' Although most of the measurement approaches that we describe were proposed for image tagging, some were proposed for other computer vision tasks but can be applied to image tagging. We cluster the approaches into five distinct~categories based on their shared analytic foundations.

\textbf{Incidence-based approaches} test whether pre-identified problematic tags, types of images, or pairings of tags and types of images occur in the inputs and outputs of image tagging systems or in the datasets used to train them. These approaches are premised on the belief that there are particular tags, types of images, and pairings that are inherently problematic: problematic tags should not be applied to any images, problematic types of images should not have any tags applied to them, and problematic pairings should not occur.\looseness=-1

For example, researchers have used incidence-based approaches to study tags that refer to properties that cannot be determined from visual appearance \cite{yang2020fairerdatasets}, such as gender \cite{keyes2018misgendering, 10.1145/3359246, 10.1145/3392866}, character traits \cite{barlas2019eye}, emotion \cite{bard2020emotion, kyriakou2019proprietary, barlas2019eye, sedenberg2017smile}, and physical attractiveness \cite{kyriakou2019proprietary}. As another example, researchers have uncovered the presence of lewd tags and images in several widely-used computer vision datasets \cite{crawford2019excavating, yang2020fairerdatasets, prabhu2020large}. Finally, one of the most widely discussed problematic pairings arose when Google Photos applied the tag \emph{gorillas} to an image of Black people \cite{simonite2018comes}. Although neither the tag nor the image would have been considered problematic in isolation, 
they become so when paired together.


Because problematic tags, types of images, or pairings must be pre-identified, using an incidence-based approach means first answering the question of \emph{which} tags, types of images, or pairings should be considered problematic---a task that requires extensive background knowledge. Yet even with such knowledge, pre-identifying \emph{all} problematic~tags,~types of images, or pairings is likely impossible.



\textbf{Distribution-based approaches} compare differences in the distributions of tags or types of images in the inputs and outputs of image tagging systems or in the datasets used to train them. One distribution-based approach compares the distribution of images that depict different social groups in a training dataset to some reference distribution, typically the uniform distribution~\cite{buolamwini2018gender, karkkainen2021fairface, wang2021revise, yang2020fairerdatasets}. A more ambitious variant of this approach focuses on the distributions of images that depict attributes (e.g., payot), artifacts (e.g., the Torah), or activities (e.g., lighting the menorah) that are bound up with the identities of different social groups, again comparing them to some reference distribution
\cite{shankar2017classification, wang2021revise, devries2019does}. Both variants are premised on the idea that image tagging systems may under-perform for social groups, attributes, artifacts, or activities that are not sufficiently well represented in the datasets used to train them. 

Another distribution-based approach compares the distributions of tags---either in a training dataset \cite{barlas2019eye, otterbacher2019talk, wang2021revise, yang2020fairerdatasets} or output by an image tagging system \cite{alvi2018removal, kyriakou2019proprietary, barlas2019eye, schwemmer2020diagnosing}---for images that depict different social groups. Other researchers have  instead compared the distributions of social groups depicted in images for different tags \cite{10.1145/2702123.2702520}. 
Both variants of this approach are premised on the belief that it is harmful for images of different social groups to have different distributions of tags, even if these tags are correctly applied, because this casts the groups in different lights.

A final approach compares the distribution of tags applied to images that depict different social groups in a image tagging system's training dataset to the distribution of tags applied by the system to images that depict these social groups \cite{wang2019balanced, wang2020fairness, zhao2017men, 10.1145/2702123.2702520, wang2021directional}. This approach is premised on the belief that the outputs of image tagging systems should not exacerbate any differences between social groups that are present in their training datasets.\looseness=-1

\textbf{Performance-based approaches} compare the performance of image tagging systems for particular tags or types of images. These approaches are premised on the belief that image tagging systems should perform equally well for different social groups. One widely discussed performance-based approach tests whether image tagging systems exhibit comparable performance for images that depict different social groups when applying tags that indicate membership in those social groups. For example, researchers have investigated whether binary gender classification systems perform better for images that depict men than they do for images that depict women \cite{bhargava2019exposing, buolamwini2018gender, hendricks2018women}. Other researchers have conducted similar analyses focusing on race, ethnicity, and age \cite{das2018mitigating, karkkainen2021fairface, kyriakou2019proprietary}. Some researchers have also taken an intersectional lens by focusing~on different combinations of social groups \cite{buolamwini2018gender, das2018mitigating}.

Another performance-based approach tests whether image tagging systems exhibit comparable performance for images that depict different social groups when applying \emph{any} tags. For example, researchers have investigated whether image tagging systems perform better when tagging everyday objects in images that depict some social groups than they do for images that depict others, finding performance disparities between genders~\cite{bhargava2019exposing, wang2020fairness}, ages~\cite{wang2020fairness}, races~\cite{buolamwini2018gender, wilson2019predictive}, socioeconomic statuses \cite{devries2019does}, and geographic locations \cite{shankar2017classification, devries2019does}---as well as different combinations thereof \cite{buolamwini2018gender, Khiyari2016FaceVS, Klare2012face_recognition}. 

\textbf{Perturbation-based approaches} involve varying aspects of images provided as inputs to image tagging systems to see whether different tags are applied, either correctly or incorrectly. These approaches are premised on two beliefs: first, that the behaviors of image tagging systems should not reflect spurious correlations in their training datasets; and second, that particular aspects of images should not be the basis for particular differences in tagging behaviors (e.g., the appearance of a person's face should not affect the occupation with which they are tagged). As a result, they are often used to investigate whether particular aspects of images are driving particular tagging behaviors, providing an an opportunity to reflect on whether these behaviors are normatively concerning  \cite{wilson2019predictive, muthukumar2018unequal, Klare2012face_recognition}. For example, if removing occluded faces from images reduces performance disparities between social groups, these disparities are likely due to occlusions rather than something more normatively concerning \cite{wilson2019predictive}.\looseness=-1

One 
perturbation-based approach focuses specifically on invariance. This approach tests whether the tags applied by image tagging systems are unaffected by perturbations of particular aspects of images that are known to be an inappropriate basis for \emph{any} differences in tagging behaviors. For example, researchers have re-taken photos using props, such as wigs \cite{rodden2017medium, League_2021}, or used GANs to generate faces with different characteristics~\cite{denton2021image, mcduff2019generative}, testing whether there are any differences in the tags applied to these images. Other researchers have selected pairs of semantically similar images that depict different social groups (e.g., similarly composed photos of CEOs of different genders), again testing for any differences in tagging behaviors \cite{otterbacher2018stereotypes, 10.1145/3442188.3445932, stock2018convnets}.

\textbf{Internals-based approaches} directly investigate the latent representations learned by image tagging systems. Like perturbation-based approaches, these approaches are premised on the belief that the behaviors of image tagging systems should not reflect spurious correlations in their training datasets and the belief that particular aspects of images should not be the basis for particular differences in tagging behaviors. One internals-based approach relies on saliency maps, which highlight the regions of images that are most salient when applying particular tags and are typically generated using gradients. This approach can therefore shed light on the reasoning behind particular tagging behaviors---for example, whether an image tagging system is focusing on the people depicted in an image when tagging objects---regardless of whether the tags in question are correctly applied. Saliency maps have been used to investigate tagging behaviors from quantitative perspectives (e.g., when tagging an object depicted in an image, investigating how much of the object's bounding box overlaps with the most salient region of the image) and qualitative perspectives (e.g., investigating what is depicted in the most salient region of an image) \cite{hendricks2018women, jia2018right, muthukumar2018unequal, singh2020dont}.

A second internals-based approach tests whether image tagging systems that are not intended to apply tags that indicate membership in social groups still encode notions of social groups in their feature representations. This suggests that they are inappropriately using these notions as the basis for tagging behaviors \cite{alvi2018removal, roberts2018implicit, song2020overlearning, wang2019balanced}.\looseness=-1

\section{Representational Harms}
\label{sec:harm_taxonomy}

Although the computational measurement approaches that we described in the previous section have played an important role in raising awareness of and developing mitigations for some of the harms caused by image tagging systems, their 
normative motivations are often unclear. These approaches are all premised on beliefs that particular properties or behaviors of image tagging systems or the datasets used to train them are self-evidently problematic, even though these properties or behaviors are rarely connected to concrete harms. This lack of clarity is not unique to image tagging systems or even computer vision systems---\citet{blodgett2020language} found that  computational approaches for measuring the ``fairness'' of natural language processing systems~often have similarly unclear normative motivations. 

In this section, we draw on the work of \citet{blodgett2020language} to argue that image tagging systems cause representational harms by contributing to the reproduction of harmful social hierarchies---that is, by casting some social groups as inferior to others and confining them to subordinate positions within society. We emphasize that representational harms are not harms to specific individuals; they relate to the standings of particular social groups, affecting \emph{all} members of those groups \cite{selbst2023unfair}. We identify four types of representational harms that can be caused by image tagging
systems, delineating how each one reproduces harmful social hierarchies. Having done this, we then introduce an additional normative concern---that is, when image tagging systems deny people the ability to self-identify---and discuss how it can give rise to representational harms. 

We do not claim to provide an exhaustive list of all of the types of representational harms that can be caused by image tagging systems. The range of possible variation for tags and images make them remarkably expressive. As a result, anticipating every way in which these variations could reproduce harmful social hierarchies is impossible. Rather, our aim is to bring greater structure precision to discussions of some of the harms
caused by image tagging systems and the many computational approaches that have been proposed to measure them. We leave open the possibility that future work will identify additional types of representational harms as our collective understanding of image tagging evolves.\looseness=-1

\subsection{Four Types of Representational Harms}

\textbf{Reifying social groups.} Social groups are socially constructed, so their boundaries are always contested and changing. Despite this inherent fluidity, social groups are often treated as if they are natural, fixed, or objective \cite{hanna2020towards}, thereby reifying beliefs about their salience and immutability and beliefs about the boundaries between them. Because image tagging systems rely on visual characteristics and are intended to apply pre-identified tags, they are especially likely to reify social groups, as well as the idea that there are relationships between social groups and visual appearance \cite{keyes2018misgendering}. For example, binary gender classification systems rely on the following beliefs: first, that gender can be determined from visual appearance; and second, that there are only two gender identities. However, relationships between gender and visual appearance are contested and unstable, as are gender identities themselves.

\textbf{Stereotyping social groups.} Image tagging systems can also perpetuate stereotypes---over-generalized beliefs about social groups that reproduce harmful social hierarchies \cite{barlas2021see, bhargava2019exposing}. For example, researchers have demonstrated that some image tagging systems systematically apply the tag \textit{nurse} to images of female doctors. Other researchers have shown that images of female snowboarders are often incorrectly tagged as depicting men \cite{hendricks2018women, zhao2017men}. In these examples, the latent representations of doctors and snowboarders learned by the image tagging systems are inextricably linked with gender---that is, to be a female medical professional is~to be a nurse, while to be a snowboarder is to be a man. 

These examples suggest that there are two ways that image tagging systems can perpetuate stereotypes. First, if an image tagging system is intended to apply tags that indicate membership in social groups, then it will necessarily associate particular visual characteristics (e.g., riding a snowboard) with particular social groups (e.g., men). This means that images that depict people with those visual characteristics will be more likely to be tagged as depicting members of those social groups. Second, if an image tagging system is intended to apply other tags, then it may associate visual characteristics that are common among members of particular social groups (e.g., long hair among women) with other properties (e.g., being a nurse). This means that tags that refer to those properties are more likely to be applied to images that depict people with those visual characteristics.\looseness=-1

Finally, image tagging systems can also perpetuate stereotypes in less overt ways. For example, some image tagging systems tend to apply tags that refer to professions to images that depict men, while applying tags that refer to appearance to images that depict women \cite{schwemmer2020diagnosing}. Even if these tags are correctly applied, they are harmful because they cast men and women indifferent lights by perpetuating the stereotype that professions are particularly salient for men, while appearance is particularly salient for women.

\textbf{Demeaning social groups.} Image tagging systems can  also demean social groups---that is, suggest that members of those groups are less worthy of respect than members of other groups.  This can occur in very overt ways, such as when image tagging systems apply tags that contain racial epithets---the very purpose of which is to belittle members of particular racial groups---or tags that are not inherently problematic but are demeaning when applied to images that depict particular social groups. As an example of the latter, applying the tag \textit{gorilla} to an image of Black people is demeaning because there is a long history of diminishing and dehumanizing Black people by likening them to gorillas \cite{simonite2018comes}. Image tagging systems can also demean particular social groups by applying tags that devalue attributes, artifacts, or activities that are bound up with the identities of those groups. For example, applying the tag \textit{costume} to an image that depicts a religious group's wedding clothing trivializes the group's traditions \cite{shankar2017no}. Evaluative terms can also be demeaning because they can be applied more (or less) often to images that depict some social groups than to images that depict others. For example, image tagging systems that apply the tag \textit{beautiful} to fewer images that depict a particular social group than they do to images that depict others cast that group as being worthy of less esteem \cite{levin2016beauty} and reinforce harmful beliefs about visual characteristics that are considered beautiful.\looseness=-1

\textbf{Erasing social groups.} Erasure occurs when image tagging systems fail to correctly apply tags that indicate membership in social groups or tags that refer to attributes, artifacts, or activities that are bound up with the identities of social groups \cite{keyes2018misgendering, buolamwini2018gender, benjamin2019race}. Erasure of a particular social group implies that members of that group  are not worthy of recognition and contributes to their further marginalization within society. For example, image tagging systems that do not apply the tag \textit{person} to images that depict people wearing hijabs erase Muslims. Similarly, image tagging systems that fail to apply the tag \textit{menorah} to images that depict menorahs---perhaps because the tag is not available, despite the availability of other tags that refer to Christian artifacts---erase Judaism.

Image tagging systems can also cause harms when they fail to acknowledge the relevance of people's membership in particular social groups to what is depicted in images---for example, by applying the tags \emph{people}, \emph{walking}, and \emph{street} to an image depicting women suffragists marching. As well as down-playing the gravity of what is depicted, these tags fail to acknowledge the injustices that woman have suffered specifically due to their gender. Erasure of this sort denies the role that membership in particular social groups plays in people's lived experiences, especially experiences relating to oppression and violence, which are often perpetrated on the basis of people's membership in particular social groups.


\subsection{Denying People the Ability to Self-Identify}
\label{sec:deprive}

In this section, we introduce an additional normative concern posed by image tagging systems---denying people the ability to self-identify---and discuss how it can give rise to representational harms. Image tagging systems are often used to sort people into social groups based on their appearance (e.g., gender classification systems), imposing tags on them without their awareness, involvement, or consent. Given the degree to which membership in social groups shapes the way that people are treated within society, including others' expectations of them \cite{10.1145/3351095.3375687}, this imposition can have profound consequences and may cause people to feel like they have been deprived of~a~crucial aspect of their autonomy \cite{hanna2020towards}.

We consider denying people the ability to self-identify to be a type of individual harm---that is, a harm that affects particular individuals as opposed to the standings of particular social groups. However, patterns in the ways that image tagging systems deny people the ability to self-identify can cause each of the four types of representational harms that we described above, as well as allocational harms. First, it reifies the belief that social groups can be determined
from visual appearance. Second, the visual characteristics that image tagging systems rely on when tagging people as 
members of social groups can perpetuate stereotypes. Third, imposing tags that indicate membership in social groups onto people can be demeaning---for example by misgendering them or by accurately applying tags (e.g., the tag \textit{women}) to some social groups (e.g., white women) but not others (e.g., Black women). Fourth, if the available tags exclude particular social groups (e.g., people who are non-binary), those groups will be erased. Finally, imposing tags that indicate membership in social groups onto people may increase their visibility and thus their vulnerability to violence, surveillance, loss of employment, and other harms.\looseness=-1

\section{Mapping Harms to Measurement Approaches}
\label{sec:mapping}

\begin{figure}[t!]
\centering
\includegraphics[width=0.45\textwidth]{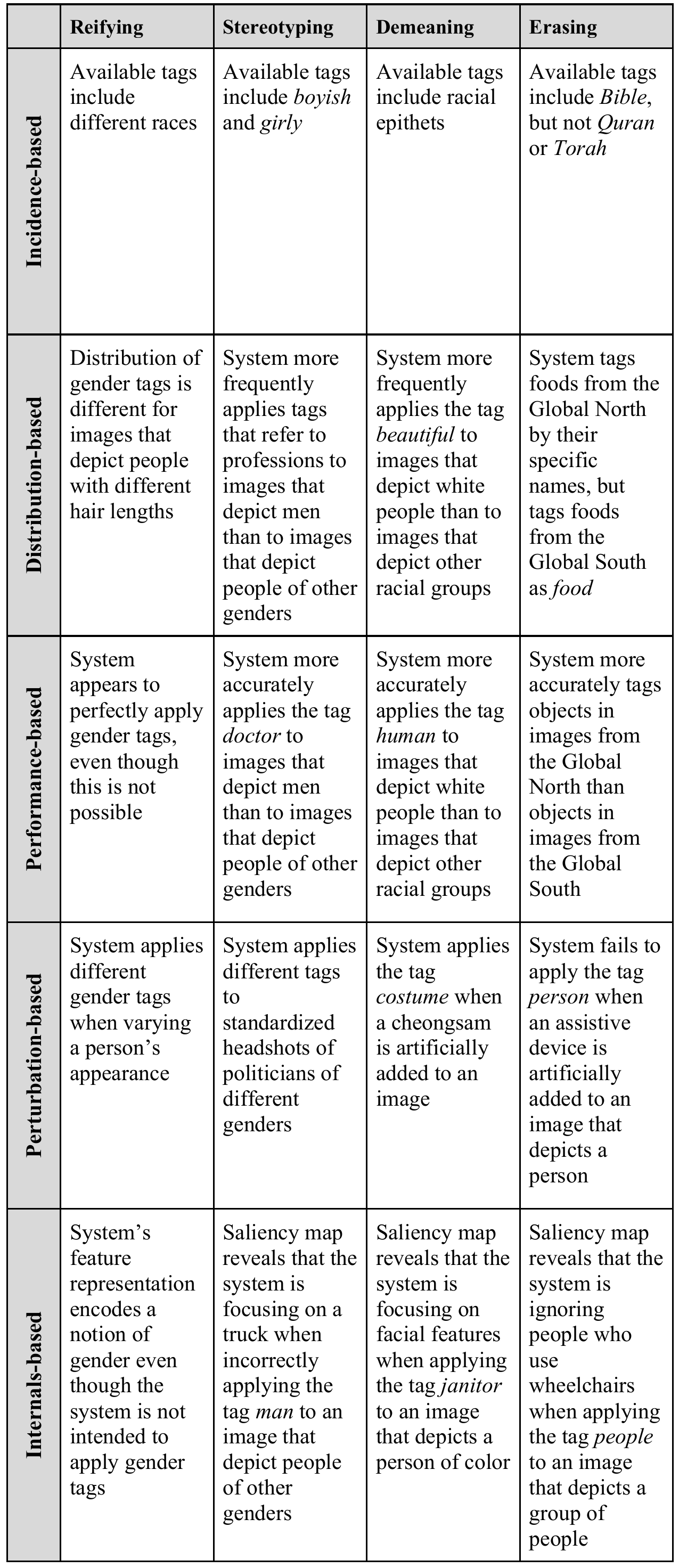}
\caption{Examples of how computational measurement approaches belonging to each of the five categories described earlier might be used to measure each of the four different types of representational harms that can be caused by image tagging systems. This mapping is by no means exhaustive.}
\label{tab:mapping}
\end{figure}

In this section, we consider how computational measurement approaches belonging to each of the five categories described earlier map to each of the four types of representational harms identified in the previous section, demonstrating that there is not
a one-to-one mapping. Due to space constraints, we only
provide an in-depth discussion for stereotyping, but include examples for the other types in figure~\ref{tab:mapping}.

\subsection{Stereotyping Social Groups}
Stereotyping can be measured using many different computational approaches. One possibility is to use an incidence-based approach to test, for example, whether pre-identified tags that contain stereotyping language, such as \emph{boyish} or \emph{girly}, are applied to any images. Another possibility is to use a distribution-based approach to see whether any differences in the distributions of tags for images that depict different social groups correspond to stereotypes---for example, whether tags that refer to professions are more frequently applied to images that depict men, while tags that refer to appearance are more frequently applied to women~\cite{schwemmer2020diagnosing}. A third possibility is to use a performance-based approach to compare, for example, the accuracy with which the tag \emph{doctor} is applied to images that depict men to the accuracy with which it is applied to images that depict people of other genders. A fourth possibility is to use a perturbation-based approach to test for invariance in the tags applied to pairs of images that are semantically equivalent, such as similarly composed photos of CEOs of different genders. A final possibility is to use an internals-based approach to investigate whether the regions of an image that are most salient when, for example, incorrectly applying the tag \emph{man} to an image that depicts a woman correspond to stereotypes about gender presentation.\looseness=-1

\subsection{Implications of these Mappings}

As demonstrated by the exercise of mapping different computational measurement approaches to different types of representational harms, no single measurement approach is
definitive. Different facets of each type of harm will be revealed by different approaches \cite{jacobs2021measurement}. As a result, measuring representational harms does not involve identifying and then using the single best measurement approach. Rather, many different approaches should be used, with the resulting measurements interpreted in the context of what those 
approaches are capable of revealing. The exercise also demonstrates that without explicit reasoning, it is not possible to infer from the use of a particular measurement approach which type of harm was intended to be measured. We therefore recommend explicitly stating the type of harm of interest so as to avoid any ambiguity or confusion. In concurrent work, we provide a concrete demonstration of how to go about this type of analysis, using a variety of computational approaches to measure different types of representational harms that can be caused by image captioning systems \cite{wang2022representational}.\looseness=-1

\section{Tensions When Mitigating Harms}

Lastly, equipped with a more granular understanding of the
types of representational harms that can be caused by image
tagging systems, we explore tensions between attempts to mitigate some of these types of harms---and what
these tensions might mean for developing appropriate mitigations.\looseness=-1


The most obvious way to mitigate reifying social groups is simply to remove tags that indicate membership in social groups. By removing such tags, image tagging systems can avoid enforcing boundaries between social groups and allow members of social groups remain unmarked, decreasing their visibility and thus their vulnerability to
violence, surveillance, loss of employment, and other harms \cite{hofmann2020living}. In contrast, the most obvious way to mitigate erasing social groups is make sure that tags that indicate membership in social groups exist and are correctly applied. In some cases, marking social groups---that is, calling them out by name---can bring much needed recognition of the injustices they have suffered. In other cases, marking privileged social groups that often go unmarked (e.g., men)~can~provide a way to draw attention to their privilege.

As another example, removing tags that refer to attributes, artifacts, or activities that are bound up with the identities of social groups can mitigate concerns that such tags might be applied incorrectly, thereby demeaning particular social groups. However, removing tags associated with particular social groups can instead cause those groups to be erased.\looseness=-1

In some scenarios, particular types of images are deemed to be inherently problematic---for example, pornographic images that stereotype and demean women---meaning that they should not have any tags applied to them. However, such images cannot be ignored or removed unless they are first identified, usually by some computer vision system explicitly trained for that purpose. Moreover, systems for ignoring or removing particular types of images can erase social groups by incorrectly labeling images that are associated with those groups as problematic. For example, when Tumblr banned ``adult content,'' they also ended up removing posts from queer communities and sex-positive communities~\cite{pilipets2020tumblr, bronstein2020tumblr}.\looseness=-1

As a final example, differences in the distributions of tags for images that depict different social groups often correspond to stereotypes. One straightforward way to mitigate these stereotypes is to ensure that the distributions of tags for different social groups are similar---for example, that the distribution of tags applied to images of Black football players is similar to the distribution of tags applied to white football players \cite{rhue2018racial}. However, constraining the distributions of tags for different social groups can be seen as a denial of the role that membership in particular social groups plays in people's lived experiences---a form of erasure.\looseness=-1


These tensions suggest that developing appropriate mitigations is a complex task, with many considerations. However, in many cases, the scenario in which an image tagging system is intended to be used can (at least partially) determine the appropriateness of different mitigations. For example, the developers of an image tagging system for a social media platform might remove particular tags to avoid stereotyping or demeaning particular social groups. In contrast, the developers of an image tagging system for blind and low-vision users might decide that the consequences of erasing social groups are worse than the consequences of incorrectly~applying tags \cite{bennett2021s, hanley2021computer}.\looseness=-1

\section{Conclusion}

In this paper, we argued that generic concerns about ``unfairness,'' ``bias,'' or ``discrimination'' in the context of image tagging should instead be understood in terms of the reproduction of harmful social hierarchies---that is, representational harms. We analyzed previously proposed computational approaches for measuring the ''fairness'' of image tagging systems, finding that they cluster into five distinct categories, each with its own analytic foundation. We also identified four types of representational harms that can be caused by image tagging
systems, delineating how each one reproduces
harmful social hierarchies. By considering how different computational measurement approaches map to each of these types of harms, we demonstrated that each approach can be used to measure all four types, with different facets of each type revealed by different approaches. Lastly, we showed that attempts to mitigate different types of representational harms may be in tension with one another and explored what these tensions might mean for developing appropriate mitigations. Moving forward, we hope that this paper contributes to more precise and actionable discussions of some of the harms caused by image tagging systems.


\bibliography{references}

\end{document}